%% file: main.tex
\documentclass[prodmode,acmtecs]{acmsmall}

\usepackage{graphicx}
\usepackage{multirow,tabularx}
\usepackage{array}
\usepackage{varwidth}
\usepackage{listings}
\usepackage{multirow}
\usepackage{url}
\usepackage{xspace}
\usepackage{amsfonts}
\usepackage{amsmath}
\usepackage{color}
\usepackage{booktabs}
\usepackage{xspace}
\usepackage{float}
\usepackage{rotating}

\newfloat{listing}{thp}{lop}
 \floatname{listing}{Listing}
 
\usepackage{cite}

\newcolumntype{Y}{>{\raggedleft\arraybackslash}X}
\newcolumntype{Z}{>{\centering\arraybackslash}X}

\newcommand{\JSTaaS}{\textsc{JSTaaS}\xspace}
\newcommand{\cs}{\textsc{JCloudScale}\xspace}

\begin{document}

\title{JCloudScale: Closing the Gap Between IaaS and PaaS}		


\author{
Rostyslav Zabolotnyi 
\affil{Distributed Systems Group, Vienna University of Technology}
Philipp Leitner
\affil{Software Evolution \& Architecture Lab, University of Zurich}
Waldemar Hummer
\affil{Distributed Systems Group, Vienna University of Technology}
Schahram Dustdar
\affil{Distributed Systems Group, Vienna University of Technology}
}

\begin{abstract}The Infrastructure-as-a-Service (IaaS) model of cloud computing
is a promising approach towards building elastically scaling systems.
Unfortunately, building such applications today is a complex, repetitive and
error-prone endeavor, as IaaS does not provide any abstraction on top of naked
virtual machines. Hence, all functionality related to elasticity needs to be
implemented anew for each application. In this paper, we present \cs, a
Java-based middleware that supports building elastic applications on top of a
public or private IaaS cloud. \cs allows to easily bring applications to the
cloud, with minimal changes to the application code.
We discuss the general architecture of the middleware as well as its technical
features, and evaluate our system with regard to both, user acceptance (based on
a user study) and performance overhead. Our results indicate that
\cs indeed allowed many participants to build IaaS applications more efficiently,
comparable to the convenience features provided by industrial Platform-as-a-Service
(PaaS)
solutions. However, unlike PaaS, using \cs does not lead to a loss of control and
vendor lock-in for the developer. 
\end{abstract}

\category{D.2.2.c}{Software Engineering}{Distributed/Internet based software
engineering tools and techniques}
\category{D.2.0.c}{Software Engineering}{Software Engineering for Internet projects}

\terms{Languages, Experimentation, Performance}



\begin{bottomstuff}
The research leading to these results has received funding from the European Community's Seventh Framework Programme (FP7/2007-2013) under grant agreement no. 610802 (CloudWave) and no. 318201 (SIMPLI-CITY).

Author's addresses: R. Zabolotnyi, W. Hummer, S. Dustdar, Institute of Information Systems 184/1, Distributed Systems Group, Vienna University of Technology, Argentinierstrasse 8, A-1040 Wien, Austria.
P. Leitner: s.e.a.l. - software evolution \& architecture lab, University of Zurich, Binzm\"uhlestrasse 14, 8050 Zurich, Switzerland.
\end{bottomstuff}

\maketitle

\input{introduction}

\input{cloudscale}
\input{elasticity}
\input{implementation}
\input{survey}
\input{relatedwork}
\input{conclusions}


\bibliographystyle{ACM-Reference-Format-Journals}
\bibliography{bibtex}

\end{document}

%% file: introduction.tex
\section{Introduction}
\label{sec:intro}

In recent years, the cloud computing paradigm~\cite{buyya:09,grossman:09} has provoked a
significant push towards more flexible provisioning of IT resources, including computing power, storage and networking capabilities.
Besides economic factors (e.g., pay-as-you-go pricing),
the core driver behind this cloud computing hype is the idea of elastic computing. Elastic applications are able to increase and decrease their resource usage based on current application load, for instance by adding and removing computing nodes. Optimally, elastic applications are cost and energy efficient (by virtue of operating close to optimal resource utilization levels), while still providing the expected level of application performance.

Elastic applications are typically built using either the IaaS (Infra\-structure-as-a-Service) or the PaaS (Platform\--as\--a\--Ser\-vice) paradigm~\cite{armbrust:10}. In IaaS, users rent virtual machines from the cloud provider, and retain full control (e.g., administrator rights). In PaaS, the level of abstraction is higher, as the cloud provider is responsible for managing virtual resources. In theory, this allows for more efficient cloud application development, as less boilerplate code (e.g., for creating and destroying virtual machines, monitoring and load balancing, or application code distribution) is required. However, practice has shown that today's PaaS offerings (e.g., Windows
Azure, Google'
AppEngine, or Amazon's Elastic Beanstalk) come with significant disadvantages, which render this option infeasible for many developers. These problems include: (1) strong vendor lock-in~\cite{lawton:08,dillon:10}, as one is typically required to program against a proprietary API; (2) limited control over the elasticity behavior or the application (e.g., developers have very little influence on when to scale up and down); (3) no root access to the virtual servers running the actual application code; and (4) little support for building applications that do not follow the basic architectural patterns assumed by the PaaS offering~\cite{jayaram:13} (e.g., Apache Tomcat based web applications).  All in all, developers are often forced to fall back to IaaS for many use cases, despite the significant advantages that the PaaS model would promise. 

In this paper, we introduce \cs, a Java-based middleware that eases the task of building elastic applications. Similar to PaaS, \cs takes over virtual machine management, application monitoring, load balancing, and code distribution. However, given that \cs is a client-side middleware instead of a complete hosting environment, developers retain full control over the behavior of their application. Furthermore, \cs supports a wide range of different applications. \cs applications run on top of any IaaS cloud, making \cs a viable solution to implement applications for private or hybrid cloud settings~\cite{sotomayor:09,abraham:10}. In summary, we claim that the \cs model is a promising compromise between  IaaS and PaaS, combining many advantages of both worlds.


The main contributions of this paper are two-fold. Firstly, we describe the 
the \cs middleware in detail. This contribution is in extension of our initial work in~\cite{leitner:12}.
Secondly, we conducted a user study to evaluate \cs in comparison to both, existing IaaS (OpenStack and Amazon EC2)
and PaaS (Amazon Elastic Beanstalk) systems. We address
 runtime performance impact of \cs, as well as development productivity and user
acceptance. Our study results suggest that \cs increases developer productivity in comparison to pure IaaS
solutions, comparable to Elastic Beanstalk. Unlike Elastic Beanstalk, \cs is more flexible,
does not lead to vendor lock-in, and can also be used in a private or hybrid cloud environment. However,
our results also show that there still are technical issues in the current \cs prototype
that need to be addressed. Further, our results show that, in its current version, \cs indeed impacts performance
in a small but noticable manner. \cs is already available as open source project from GitHub. 


The rest of this paper is structured as follows. In Section~\ref{sec:cloudscale},
we describe the basic \cs architecture, which we follow up with an in-depth discussion
of specific elasticity-related features in Section~\ref{sec:elasticity}.
Section~\ref{sec:impl}
gives an implementation overview of the middleware. This implementation forms the
basis for the empirical evaluation in Section~\ref{sec:survey}.
Section~\ref{sec:related} surveys related work, and, finally,
Section~\ref{sec:conclusions} concludes the paper with an outlook on open
issues.

%% file: cloudscale.tex
\section{The CloudScale Middleware}
\label{sec:cloudscale}

In the following, we introduce the main notions and features of \cs.

\subsection{Basic Notions}
\label{sec:basic}

\cs is a Java-based middleware for building elastic IaaS applications. The
ultimate aim of \cs is to facilitate developers to implement cloud applications
(in the following referred to as \emph{target applications}) as local,
multi-threaded applications, without even being aware of the cloud deployment.
That is, the target application is not aware of the underlying physical
distribution, and does not need to care about technicalities of elasticity, such
as program code distribution, virtual machine instantiation and destruction,
performance monitoring, and load balancing. This is achieved with a declarative
programming model (implemented via Java annotations) combined with bytecode
modification. To the developer, \cs appears  as an additional library (e.g., a
Maven dependency) plus a post-compilation build step. This puts \cs in stark
contrast to most industrial PaaS solutions, which require applications to be
built specifically for these platforms. Such PaaS applications are usually not
executable outside of the targeted PaaS environment.

\begin{figure}[h!]
\centering
\includegraphics[width=0.45\linewidth]{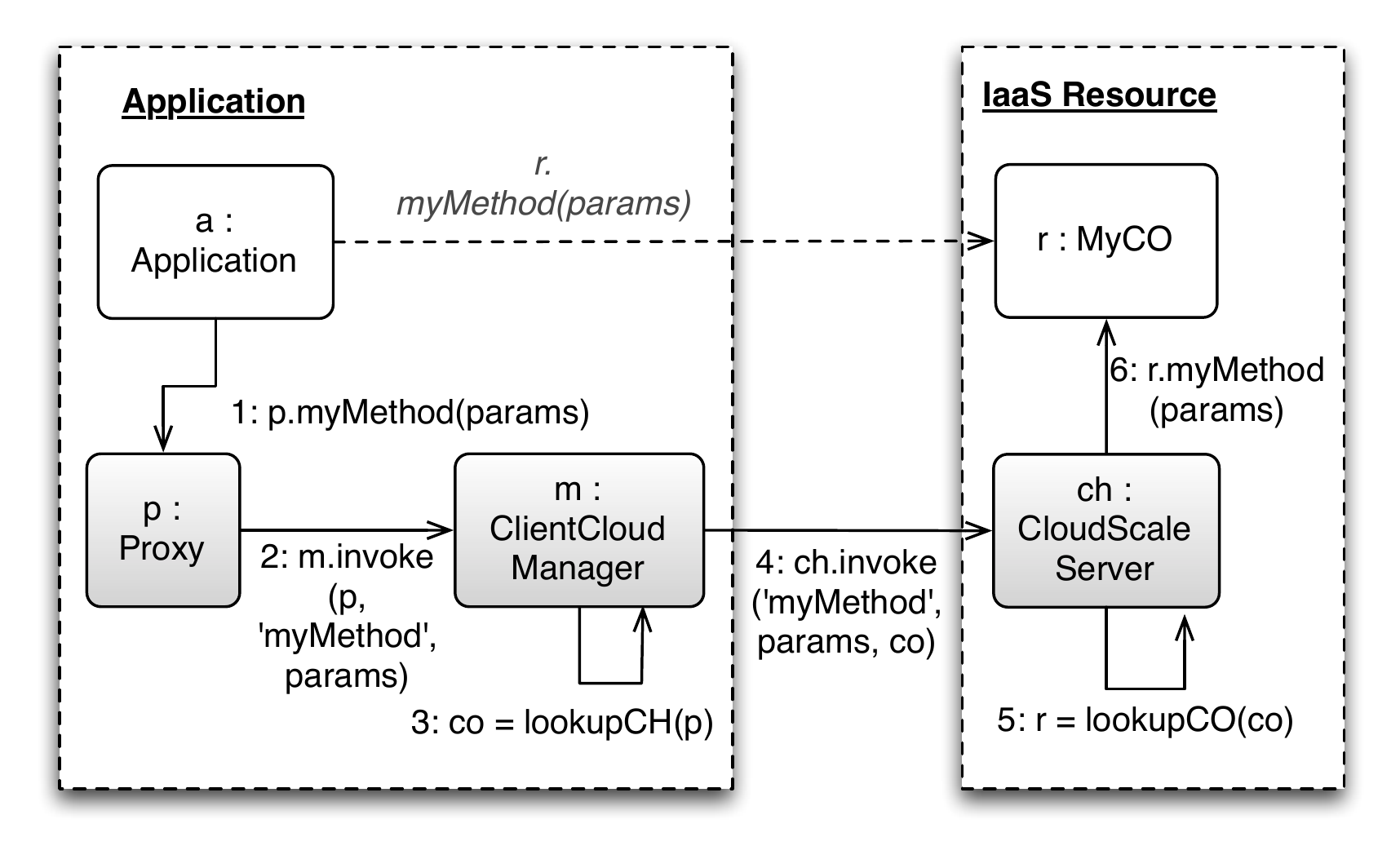}
  \caption{Basic Interaction with Cloud Objects}
  \label{fig:interacton}
\end{figure}

The primary entities of \cs are \emph{cloud objects} (COs). COs are object
instances which execute in the cloud. COs are deployed to, and executed by,
so-called \emph{cloud hosts} (CHs). CHs are virtual machines acquired from the
IaaS cloud, which run a \cs server component. They accept COs to host and
execute on client request.
The program code responsible for managing virtual machines, dispatching
requests to virtual machines, class loading, and monitoring is injected into
the target application as a post-compilation build step via bytecode
modification. Optimally, COs are highly cohesive and
loosely coupled to the rest of the target application, as, after cloud
deployment, every further interaction with the CO constitutes a remote
invocation over the network.

\begin{figure}[h!]
\centering
\includegraphics[width=0.35\linewidth]{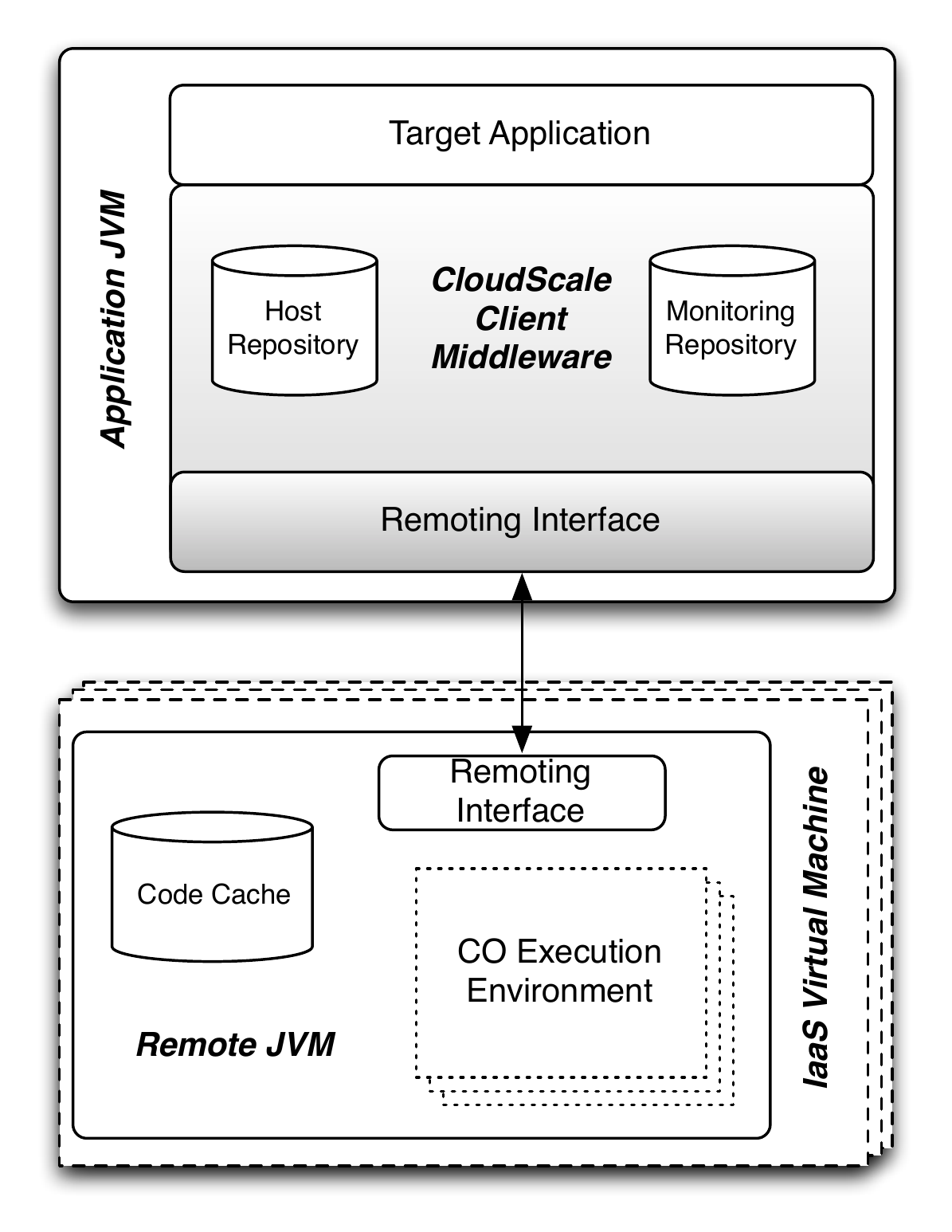}
  \caption{System Deployment View}
  \label{fig:architecture}
\end{figure}

Fig.~\ref{fig:interacton} illustrates the basic operation of
\cs in an interaction diagram. The grey boxes indicate code that is injected.
Hence, these steps are transparent to the application developer.

Fig.~\ref{fig:architecture} shows a high-level deployment view of a \cs application. The grey box in the target application JVM again indicates injected components. Note that CHs are conceptually ``thin'' components, i.e., most of the actual \cs business logics is running on client side. CHs consist mainly of a small server component that accepts requests from clients, a code cache used for classloading, and sand boxes for executing COs. As \cs currently does not explicitly target multi-tenancy~\cite{bezemer:10}, these sand boxes are currently implemented in a light-weight way via custom Java classloaders. On client-side, the \cs middleware collects and aggregates monitoring data, and maintains a list of CHs and COs. Further, the client-side middleware is responsible for scaling up and down based on user-defined policies (see Section~\ref{sec:scaling}). 

\begin{figure}[h]
\scriptsize
\lstset{float=thpb}
\begin{lstlisting}[caption=Declaring COs in Target Applications, label=code:example]
@CloudObject
public class MyCO {
  
  @CloudGlobal
  private static String myCoName; 
  
  @DataSource(name = "couchdb")
  private DataStore datastore;
  
  @EventSink
  private EventSink eventsink;
  
  public MyResult myMethod(@ByValueParameter MyParameters params) {
     ...
  }
}
\end{lstlisting}
\end{figure}

\subsection{Interacting with Cloud Objects}
\label{sec:interacting}

Application developers declare COs in their application code via simple Java
annotations (see Listing~\ref{code:example} for a minimal example).
As is the case for any object in Java, the target application can fundamentally
interact with COs in two different ways: invoking CO methods, and getting or
setting CO member fields. In both cases, \cs intercepts the operation, executes
the requested operation on the CH, and returns the result (if any) back to the
target application. In the meantime, the target application is blocked (more
concretely, the target application remains in an ``idle wait'' state while it is
waiting for the CH response). Fundamentally, \cs aims to preserve the functional
semantics of the target application after bytecode modification. That is, every
method call or field operation behaves functionally identical to a regular Java
program.   

One exception to this rule are CO-defining classes that contain static fields and  methods.
Operations on those are by default not intercepted by \cs,
as they potentially lead to a problem that we refer to as \emph{JVM-local
updates}: if code executing on a CH (for instance a CO instance method) changes
the value of a static field, only the copy in this CH's JVM will be changed.
Other COs, or the target application JVM, are not aware of the change. Hence, in
this case, the value of the static field is tainted, and the execution semantics
of the application changes after \cs bytecode injection. To prevent this
problem and preserve standard Java language semantics, static fields can be
annotated with the \texttt{@CloudGlobal}
annotation  (see Listing~\ref{code:example}, line 4-5). Changes to such static fields are
maintained in the target application JVM, and all CH JVMs are operating on the
target application JVM copy via callback. Note that this behavior is not default
for performance reasons, as synchronizing static field values is expensive, and
only required if JVM-local updates are possible.

\subsection{Remote Classloading}
\label{sec:remoteclassloading}

Whenever a CH has to execute a CO method, \cs has to ensure that all necessary
resources (i.e., program code and other files, for instance configuration files)
are available on that CH. In order to ensure freshness of the available code
and to retrieve missing files, we intercept the default class loading mechanism
of Java and verify that the code available to the CH is the same as the one
referenced by the client. If this is not the case, the correct version of the
code is fetched dynamically from the target application. In order to improve
performance, CHs additionally maintain a code cache, which is a high-speed
storage of recently used code.
This mechanism allows \cs to load missing or modified code efficiently and
seamlessly for the application only whenever it is necessary, thus simplifying
application development and maintenance. We discuss this process in more detail
in~\cite{zabolotnyi:13}.



%

%% file: elasticity.tex
\section{Supporting Cloud Elastic Applications}
\label{sec:elasticity}

So far, we have discussed how \cs transparently enables remoting in cloud
applications. We now explain how \cs enables elastic applications.

\subsection{Autonomic Elasticity via Complex Event Processing}
\label{sec:scaling}

One central advantage of \cs is that it allows for building elastic applications
by mapping requests to a dynamic pool of CHs. This encompasses three related
tasks: (1) performance monitoring, (2) CH provisioning and de-provisioning, and
(3) CO-to-CH scheduling and CO migration. One design goal of \cs is to abstract
from technicalities of these tasks, but still grant developers low-level control over
the elasticity behavior.

\begin{figure}[h!]
\centering
\includegraphics[width=0.45\linewidth]{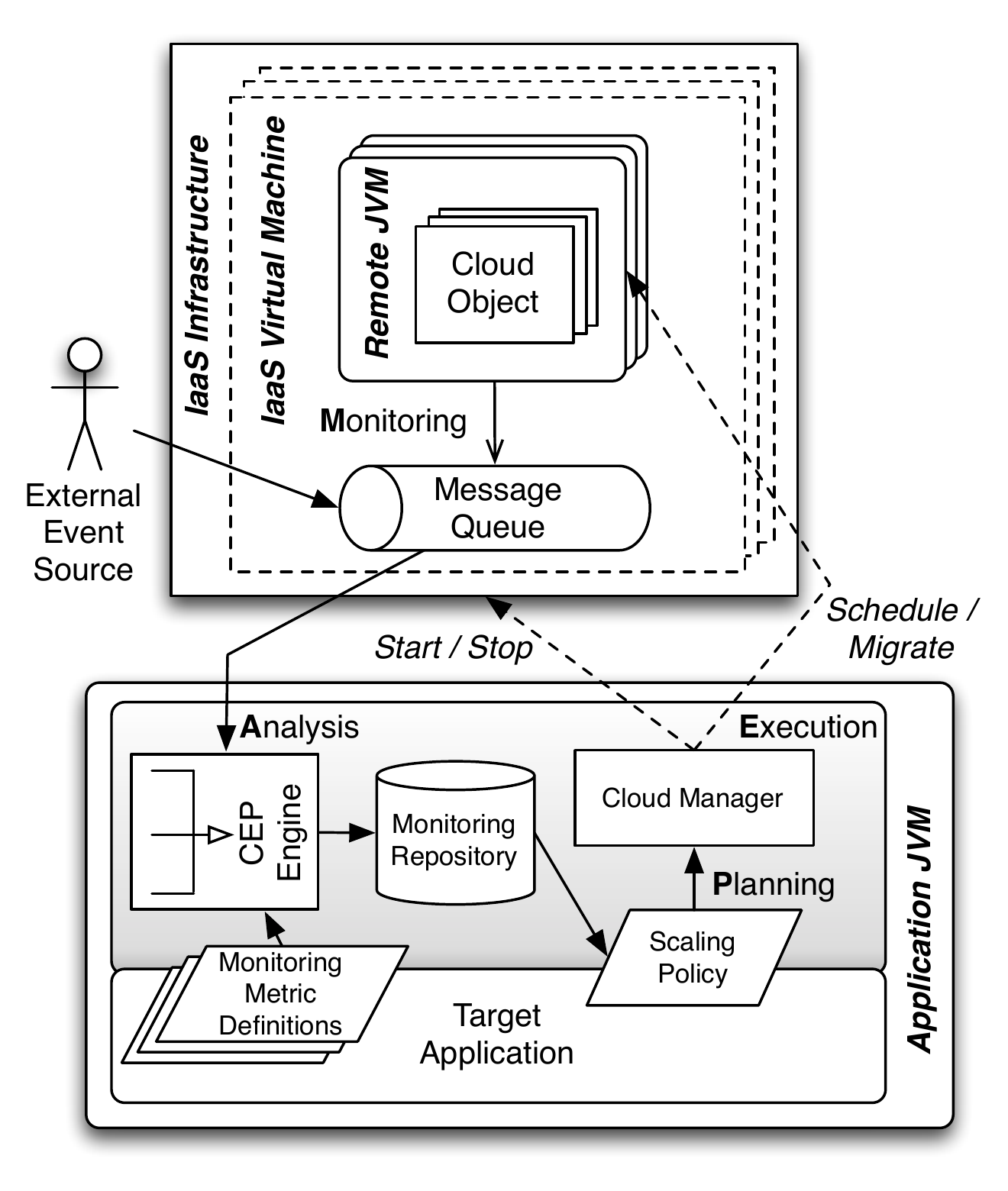}
  \caption{Autonomic Elasticity}
  \label{fig:cep}
\end{figure}

An overview over the \cs components related to elasticity, and their
interactions, is given in Fig.~\ref{fig:cep}. Conceptually, our system
implements the well-established autonomic computing control loop of
monitoring\--analysis\--plan\-ning\--execution~\cite{kephart:03} (MAPE). The
base data of monitoring is provided using event messages. All components in a
\cs system (COs, CHs, as well as the middleware itself) trigger a variety of
predefined lifecycle and status events, indicating, for instance, that a new CO
has been deployed or that the execution of a CO method has failed. Additionally,
\cs makes it easy for applications to trigger custom (application-specific)
events. Finally, events may also be produced by \emph{external event sources},
such as an external monitoring framework. All these events form a consolidated
stream of monitoring events in a \emph{message queue}, by which they are
forwarded into a \emph{complex event processing (CEP) engine}~\cite{luckham} for
analysis. CEP is the process of merging a large number of low-level events into
high-level knowledge, e.g., many atomic execution time events can be merged into
meaningful performance indicators for the system in total.
  
Developers steer the scaling behavior by defining a \emph{scaling policy}, which
implements the planning part of this MAPE loop. This policy is invoked whenever
a new CO needs to be scheduled. Possible decisions of the \emph{scaling policy}
are the provisioning of new CHs, migrating existing COs between CHs, and
scheduling the new CO to a (new or existing) CH. The policy is also responsible
for deciding whether to de-provision an existing CH at the end of each billing
time unit. Additionally, developers can define any number of \emph{monitoring
metrics}. Metrics are simple 3-tuples \emph{$<$name, type, cep-statement$>$}.
CEP-statements are defined over the stream of monitoring events. An example,
which defines a metric \texttt{AvgEngineSetupTime} of type
\texttt{java.lang.Double} as the average \texttt{duration} value of all
\texttt{EngineSetupEvent}s received in a 10 second batch, is given in
Listing~\ref{code:cep}.

\begin{figure}[h]
\lstset{float=thpb}
\begin{lstlisting}[caption=Defining Monitoring Metrics via CEP, label=code:cep]
MonitoringMetric metric =
  new MonitoringMetric();
metric.setName("AvgEngineSetupTime");
metric.setType(Double.class);
metric.setEpl(
  "select avg(duration)
   from EngineSetupEvent.win
     :time_batch(10 sec)"
);
EventCorrelationEngine.getInstance()
  .registerMetric(metric); 
\end{lstlisting}
\end{figure}

\emph{Monitoring metrics}
range from very simple and domain-independent (e.g., calculating
the average CPU utilization of all CHs) to rather application-specific ones, such as the
example given in Listing~\ref{code:cep}. 
Whenever the CEP-statement is triggered, the CEP engine writes a new value to an in-memory \emph{monitoring repository}.
\emph{Scaling policies} have access to this repository, and make use of its
content in their decisions.
In combination with monitoring metrics, scaling policies are a well-suited tool
for developers to specify how the application
should react to changes in its work load. Hence, sophisticated scaling policies that minimize
cloud infrastructure costs or that maximize utilization~\cite{genaud:11}
are easy to integrate. As part of the \cs release, we provide a small number of default
policies that users can integrate out of the box, but expect users to write
their own policy for non-trivial applications. This has proven necessary as, generally,
no generic scaling policy is able to cover the needs of all applications.

Finally, the \emph{cloud manager} component, which
can be seen as the heart of the \cs client-side middleware and the executor of
the MAPE loop, enacts the decisions of the policy by invoking the respective
functions of the IaaS API and the CH remote interfaces (e.g., provisioning of
new CHs, de-provisioning of existing ones, as well as the deployment or
migration of COs).

\begin{figure}[h!]
\centering
\includegraphics[width=0.6\linewidth]{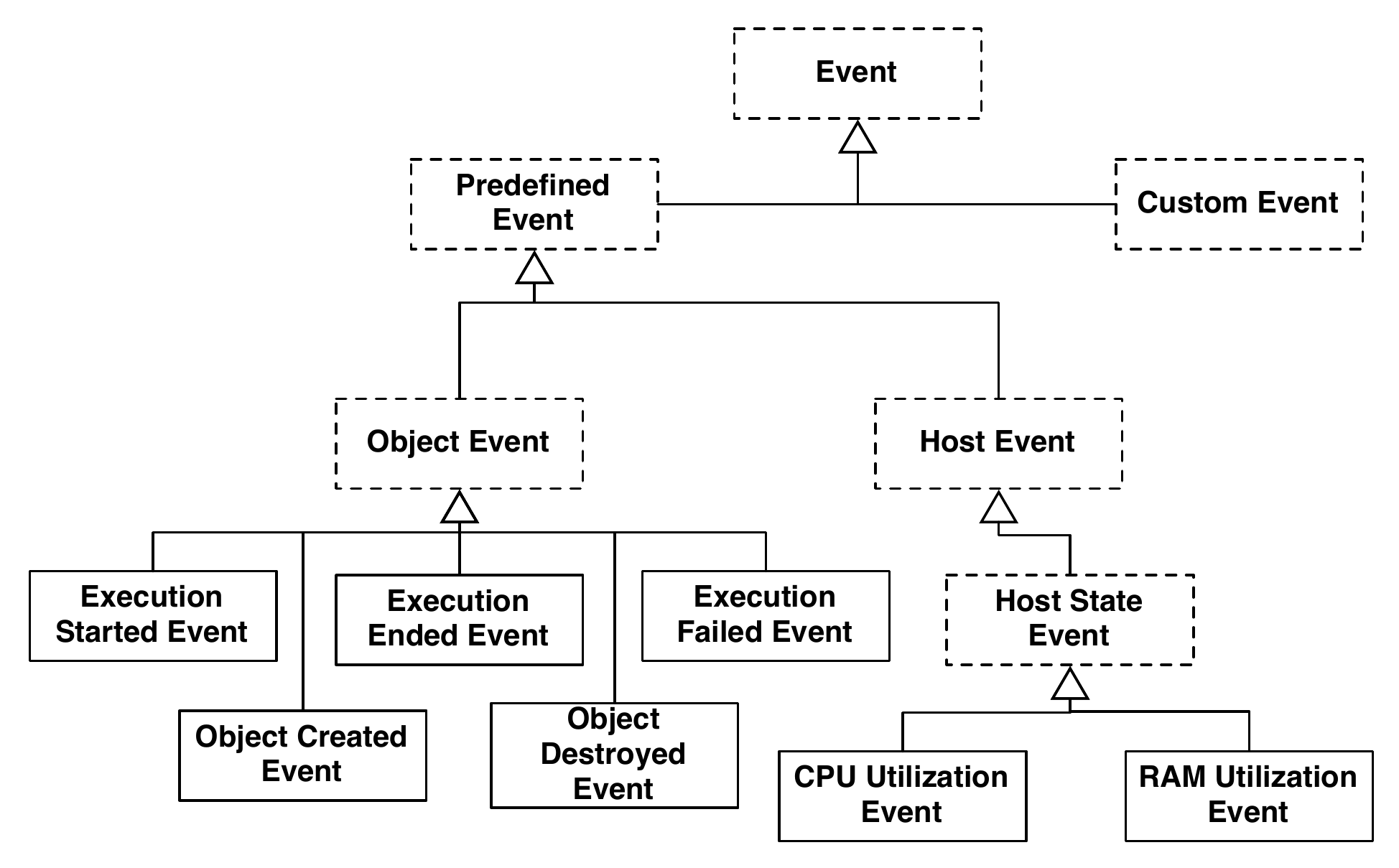}
  \caption{Monitoring Event Hierarchy}
  \label{fig:events}
\end{figure}

Fig.~\ref{fig:events} depicts the type hierarchy of all predefined events in
\cs. Dashed classes denote abstract events, which are not triggered directly,
but serve as classifications for groups of related events. All events
further contain a varying number of event properties, which form the core
information of the event. For instance, for \texttt{ExecutionFailedEvent},
the properties contain the CO, the invoked method, and the actual error.
Developers and \emph{external event sources} can extend this event hierarchy by
inheriting from \texttt{CustomEvent}, and writing these custom events into a
special event sink (injected by the middleware, see
Listing~\ref{code:example}). This process is described in more detail in~\cite{leitner:12:1}.

\subsection{Deploying to the Cloud}
\label{sec:deployment}

As all code that interacts with the IaaS cloud is injected, the \cs programming model naturally decouples Java applications from the cloud environment that they are physically deployed to. This allows developers to re-deploy the same application to a different cloud simply by changing the respective parts of the \cs configuration. \cs currently contains three separate cloud backends, supporting OpenStack-based private clouds, the Amazon EC2 public cloud, and a special \emph{local environment}. The local environment does not use an actual cloud at all, but simulates CHs by starting new JVMs on the same physical machine as the target application.  Support for more IaaS clouds, for instance Microsoft Azure's virtual machine cloud, is an ongoing activity.

\begin{figure}[h!]
\centering
\includegraphics[width=0.5\linewidth]{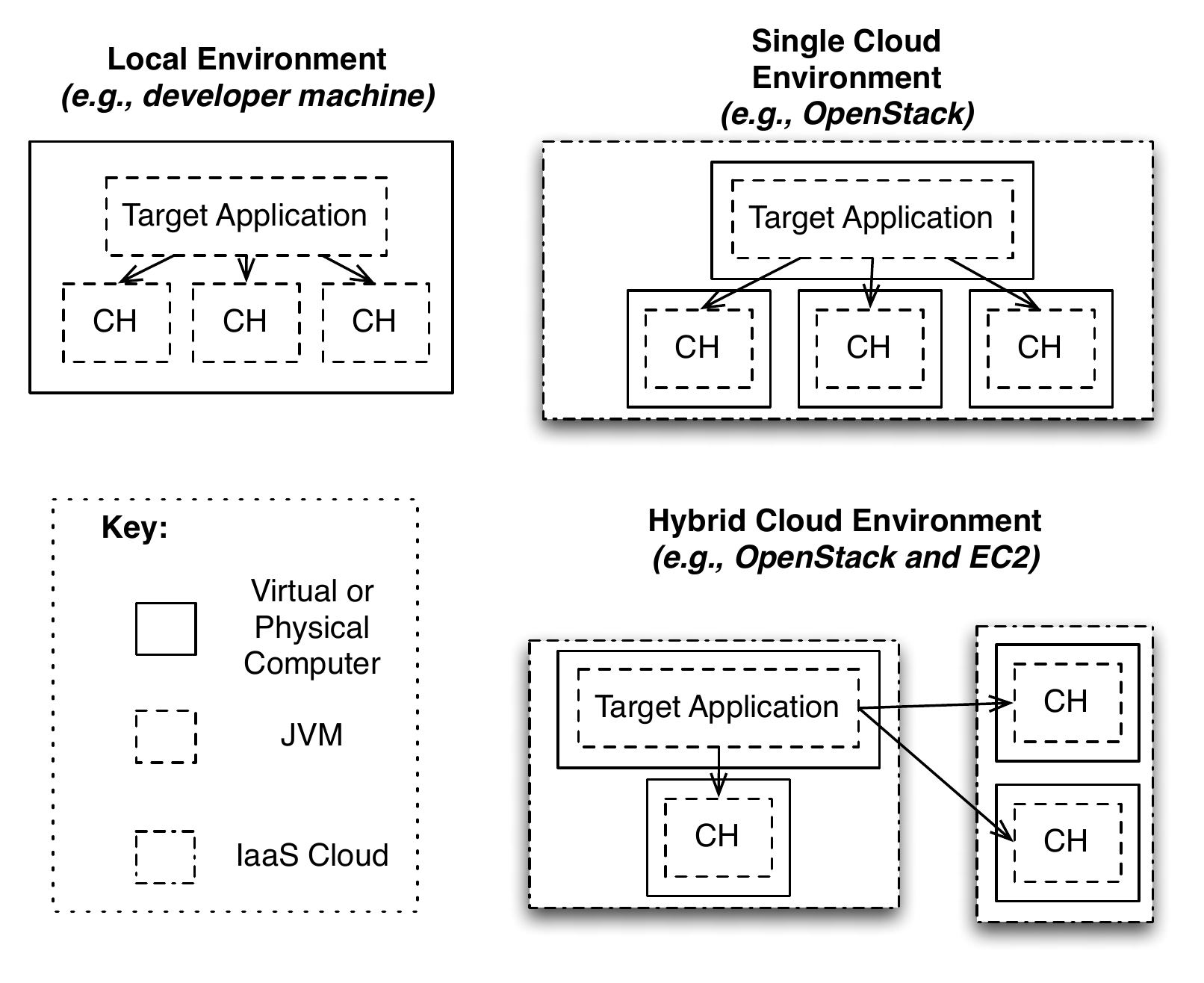}
  \caption{Supported Deployment Environments}
  \label{fig:envs}
\end{figure}

It is also possible to combine different environments, enabling hybrid cloud applications. In this case, the scaling policy is responsible for deciding which CO to execute on which cloud. Fig.~\ref{fig:envs} illustrates the different types of environments supported by \cs.

\begin{figure}[h!]
\centering
\includegraphics[width=0.35\linewidth]{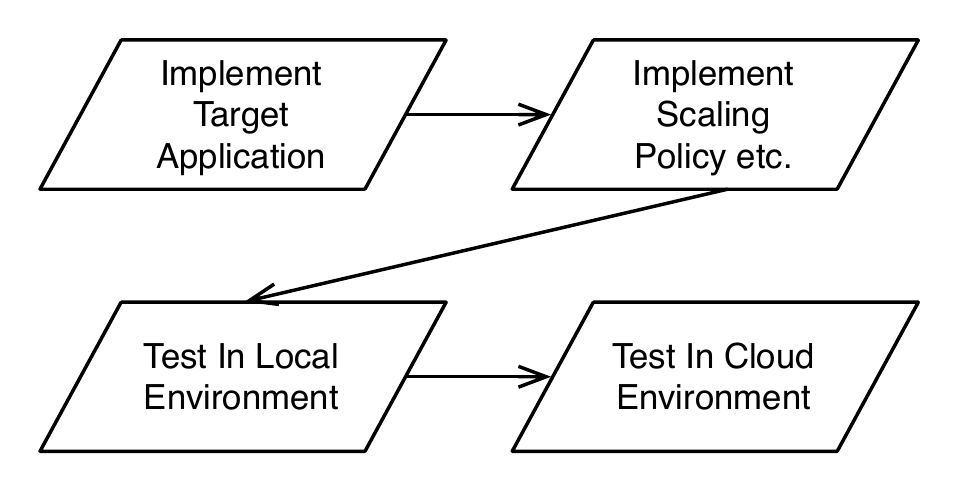}
  \caption{Conceptual Development Process}
  \label{fig:process}
\end{figure}

\subsection{Development Process}

As \cs makes it easy to switch between different cloud environments, the
middleware supports a streamlined development process for elastic applications,
as sketched in Fig.~\ref{fig:process}. The developer typically starts by
building her application as a local, multi-threaded Java application using
common software engineering tools and methodologies.
Once the target application logic is implemented and tested, she adds the
necessary \cs annotations, as well as scaling policies, monitoring metric
definitions, and \cs configuration as required. Now she enables \cs code
injection by adding the necessary post-compilation steps to the application
build process. Via configuration, the developer specifies a deployment in the
local environment first. This allows her to conveniently test and debug the
application on her development machine, including tuning and customizing the
scaling policy. Finally, once she is satisfied with how the application behaves,
she changes the configuration to an actual cloud environment, and deploys and
tests the application in a physically distributed fashion.

We argue that this process significantly lessens the pain that developers experience when building applications for IaaS clouds, as it reduces the error-prone and time-consuming testing of applications on an actual cloud. However, of course this process is idealized. Practical usage shows that developers will have to go back to a previous step in the process on occasion. For instance, after testing the scaling behavior in the local environment, the developer may want to slightly adapt the target application to better support physical distribution. Still, we have found that the conceptual separation of target application development and implementation of the scaling behavior is well-received by developers in practice. 

%% file: implementation.tex
\section{Implementation}
\label{sec:impl}

We have implemented \cs as a Java-based middleware under an Apache Licence 2.0. The current stable version is available from GitHub\footnote{https://github.com/xLeitix/jcloudscale}. This web site also contains documentation and sample applications to get users started with \cs.

\begin{figure}[h!]
\centering
\includegraphics[width=0.75\linewidth]{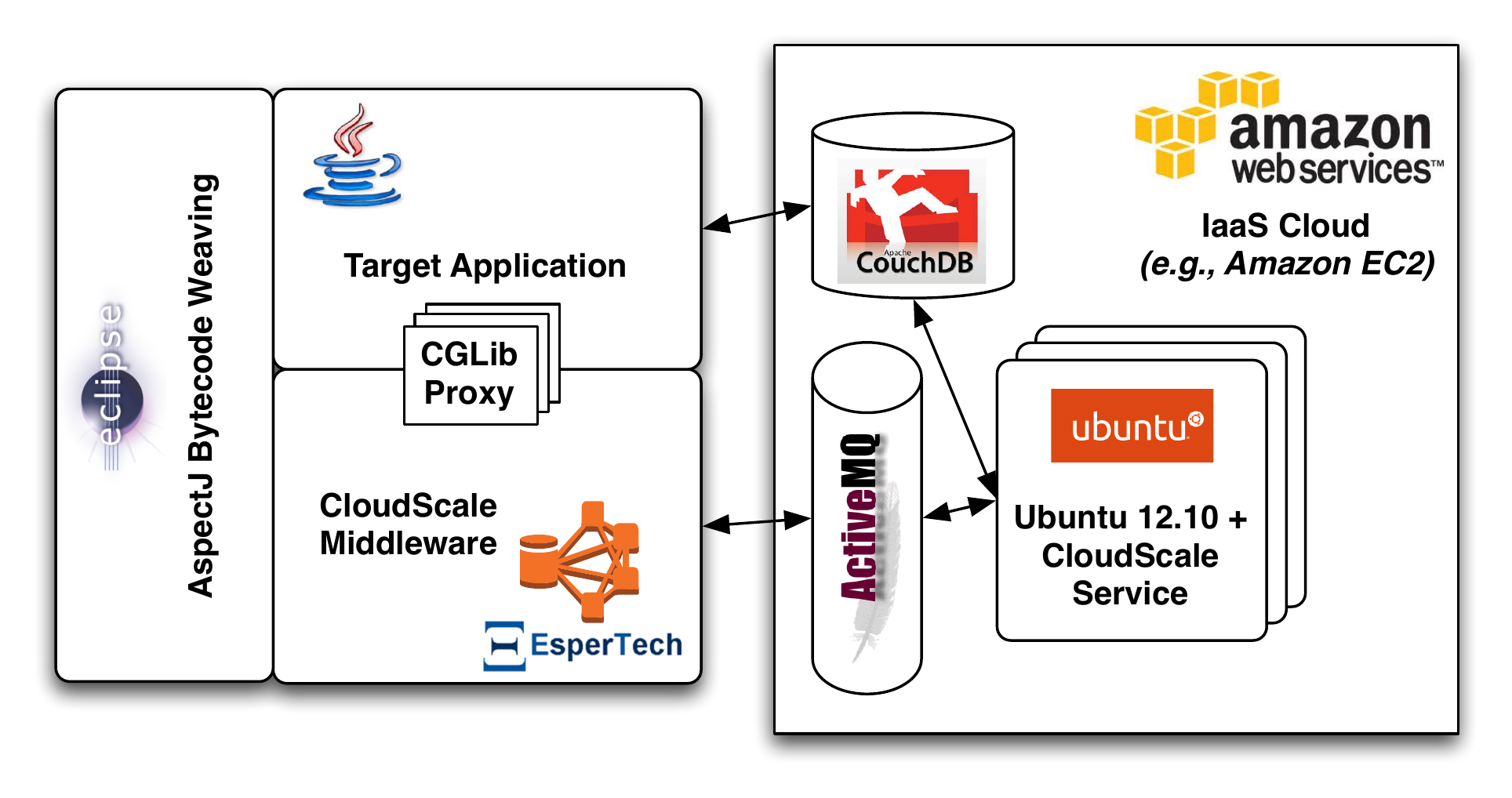}
  \caption{Implementation Overview}
  \label{fig:implementation}
\end{figure}

Our implementation integrates a plethora of existing technologies,
which is summarized in Fig.~\ref{fig:implementation}. \cs uses aspect-oriented
programming~\cite{kiczales:01} (via AspectJ) to inject remoting and cloud
management code into target applications. Typically, this is done as a
post-compilation step in the build process. Dynamic proxying is implemented by
means of the CGLib code generation library. For event processing, the
well-established Esper CEP engine is used. The client-side middleware interacts
with CHs via a JMS-compatible message queue (currently Apache ActiveMQ).
Furthermore, COs and the target application itself can read from and write to a
shared data store (for example Apache CouchDB). CHs themselves are simple Ubuntu
12.10 Linux hosts running Java and a small \cs operating system service, which
receives CO requests and executes them. Currently, we have built CH base images
for OpenStack and Amazon EC2, which are linked from the Google Code web site,
and which can be used out of the box with the stable version 0.4.0 of \cs (the
current version at the time of writing). We will also provide images for future
stable versions, once they become available.

%% file: survey.tex
\section{Validation}
\label{sec:survey}


As part of our validation of the \cs framework, we aim at answering the following three research questions:

\begin{itemize}
\item \textbf{RQ1:} Does using \cs instead of established tooling lead to more efficient development of cloud solutions, e.g., in terms of solution size or development time?
\item \textbf{RQ2:} How does \cs compare with established tooling in terms of ease-of-use, debugging, and other more ``soft'' quality dimensions?
\item \textbf{RQ3:} What runtime overhead does \cs impose at execution time? 
\end{itemize}

In order to answer \textbf{RQ1} and \textbf{RQ2}, we conducted a multi-month user study.
\textbf{RQ3} is addressed via numerical overhead measurements on a simple example application,
with and without \cs.

\subsection{User Study}
\label{sec:study}

In order to evaluate \textbf{RQ1} and \textbf{RQ2}, we
conducted an user study with 14 participants  
to assess the developers' experience with \cs as compared to using standard tools.



\subsubsection{Study Setup and Methodology}
\label{sec:study_setup}

We conducted our study with, in total, 14 male master students of computer
science at TU Vienna (participants P01 -- P14), and based on two different
non-trivial implementation tasks.
The first task was to develop a parallel computing implementation of a genetic
algorithm (\textbf{T1}). The second task required the participants to implement
a service that executes JUnit test cases on demand (\textbf{T2}). For both
tasks, an elastic solution was asked for, which was able to react to changes in
load dynamically and automatically by scaling up and down in the cloud. Both
\textbf{T1} and \textbf{T2} required roughly a developer week of effort
(assuming that the respective participant did not have any particular prior
experience with the used technologies).

The study ran in two phases. In Phase (1), we compared using \cs on top of
OpenStack with programming directly via the OpenStack API, without any specific
middleware support. This phase reflected a typical private
cloud~\cite{dillon:10} use case of \cs.
In Phase (2), we compare \cs on top of Amazon EC2 with using Amazon Elastic
Beanstalk. This reflects a common public cloud usage of the framework.
In both study phases, we asked participating developers to build solutions for
both tasks using \cs and the respective comparison technology, and compare the
developer experience based on quantitative and qualitative factors. We had 9
participating developers in Phase (1), and 5 participants in Phase (2). One
participant in Phase (2) only completed one of the two tasks.

\begin{table}[h!]
\tbl{Study Participant Overview}{
\begin{tabular}{c|c|c|c|c|c|c|c}
\textbf{ID} & \textbf{Phase} & \textbf{Java Exp.} & \textbf{Cloud Exp.} & \textbf{JCS/OS} & \textbf{OS} & \textbf{JCS/EC2} & \textbf{Beanstalk} \\
\hline
\textbf{P01} & Phase (1) & + & + & T1 & T2 & -- & --\\
\hline
\textbf{P02} & Phase (1) & + & + & T1 & T2 & -- & --\\
\hline
\textbf{P03} & Phase (1) & $\mathtt{\sim}$ & $\mathtt{\sim}$ & T2 & T1 & -- & --\\
\hline
\textbf{P04} & Phase (1) & - & - & T1 & T2 & -- & --\\
\hline
\textbf{P05} & Phase (1) & $\mathtt{\sim}$ & - & T2 & T1 & -- & --\\
\hline
\textbf{P06} & Phase (1) & + & - & T2 & T1 & -- & --\\
\hline
\textbf{P07} & Phase (1) & + & + & T2 & T1 & -- & --\\
\hline
\textbf{P08} & Phase (1) & + & $\mathtt{\sim}$ & T2 & T1 & -- & --\\
\hline
\textbf{P09} & Phase (1) & + & $\mathtt{\sim}$ & T2 & T1 & -- & --\\
\hline
\hline
\textbf{P10} & Phase (2) & + & + & -- & -- & T2 & T1\\
\hline
\textbf{P11} & Phase (2) & + & $\mathtt{\sim}$ & -- & -- & T1 & T2\\
\hline
\textbf{P12} & Phase (2) & + & - & -- & -- & T1 & T2\\
\hline
\textbf{P13} & Phase (2) & + & $\mathtt{\sim}$ & -- & -- & T2 & T1\\
\hline
\textbf{P14} & Phase (2) & + & + & -- & -- & T2 & -- \\
\hline
\end{tabular}}
\label{tab:participants}
\end{table}

Phase (1) of the study lasted two months. We initially presented \cs and the
comparison technologies to the participants, and randomly assigned which of the
tools each participant should be using for T1. Participants then had one month
of time to submit a working solution to the task along with a short report,
after which they could start working on T2 with the remaining technology. Similar to
T1, participants were given one month of time to submit a solution and a short
report. Based on the lessons leared from Phase (1), we slightly clarified and
improved the task descriptions and gave pariticpants more time (1.5 months per task)
for Phase (2). Other than that, Phase (2) was executed identically to Phase (1).  


Table~\ref{tab:participants} summarizes the relevant background for each
participant of the study. To preserve anonymity, we classify the self-reported
background of participants related to their Java or cloud experience into three
groups: relevant work experience (+), some experience ($\mathtt{\sim}$), or
close to no experience (-). The last four columns indicate whether the
participant submitted solutions for \cs running on top of OpenStack, OpenStack
directly, \cs running on top of EC2, or AWS Elastic Beanstalk, as well as which
task the participant solved using these (combinations of) technologies.

For the OpenStack-related implementations, we used a private cloud system hosted
at TU Vienna. This OpenStack instance consists of 12 dedicated Dell blade
servers with 2 Intel Xeon E5620 CPUs (2.4 GHz Quad Cores) each, and 32 GByte
RAM, running on OpenStack Folsom (release 2012.2.4). For the study, each
participant was alloted a quota of up to 8 very small instances (1 virtual CPU,
and 512 MByte of RAM), which they could use to implement and test their
solutions. For the AWS-related implementations, participants were assigned an
AWS account with sufficient credit to cover their implementation and testing
with no particular limitations.


\subsubsection{Comparison of Development Efforts (RQ1)}
\label{sec:dev_effort}

\begin{table}[h!]
\tbl{Solutions Sizes (in Lines of Code)}{
\begin{tabular}{rr|c|c|c|c|c|c|c|c}
& &  \multicolumn{4}{c|}{\textbf{Phase (1)}} & \multicolumn{4}{c}{\textbf{Phase (2)}} \\

& &  \multicolumn{2}{c|}{\textbf{JCS/OS}} & \multicolumn{2}{c|}{\textbf{OS}} & \multicolumn{2}{c|}{\textbf{JCS/EC2}} & \multicolumn{2}{c}{\textbf{Beanstalk}} \\
&   & $\tilde{A}$ & $\sigma_{A}$ & $\tilde{B}$ & $\sigma_{B}$ & $\tilde{C}$ & $\sigma_{C}$ & $\tilde{D}$ & $\sigma_{D}$ \\
    \textbf{T1} & & & & & & & & \\
  \hline
&  Business Logics & \textbf{200} & 176 & 552 & 215 & \textbf{388} & 152 & 825 & 947 \\
&  Cloud Management & \textbf{100} & 112 & 180 & 86 & \textbf{163} & 24 & 676 & 742 \\
&  Other Code 		& \textbf{170} & 157 & 286 & 226 & 1590 & 1203 & \textbf{897} & 1127 \\
&  Entire Application & \textbf{400} & 416 & 1050 & 434 & \textbf{2141} & 1331 & 2790 & 2660 \\
   \hline
    \textbf{T2} & & & & & & & & \\
  \hline
&  Business Logics & 450 & 292 & \textbf{375} & 669 & 800 & 434 & \textbf{208} & 280 \\
&  Cloud Management & \textbf{100} & 48 & 250 & 364 & \textbf{118} & 745 & 223 & 38 \\
&  Other Code 		& 325 & 213 & \textbf{300} & 297 & \textbf{140} & 2240 & 1290 & 972 \\
&  Entire Application & \textbf{1025} & 461 & 1500 & 901 & \textbf{1000} & 3328 & 2184 & 968 \\
\end{tabular}}
\label{tab:locs}
\end{table}

\textbf{RQ1} asked whether
\cs makes it easier and faster to build elastic IaaS applications. To this end,
we asked participants to report on the size of their solutions (in lines of
code, without comments and blank lines). The results are summarized in
Table~\ref{tab:locs}.
$\tilde{A}$ -- $\tilde{D}$ represent the median size of solutions, while $\sigma_{A}$
-- $\sigma_{D}$ indicate standard deviations.
It can be seen that using \cs indeed generally reduces the total source code size of
applications. Going into the study, we expected \cs to mostly reduce the amount
of code necessary for interacting with the cloud. However, our results indicate
that using \cs also often reduced the amount of code of the application
business logics, as well as assorted other code (e.g., data structures). When
investigating these results, we found that participants considered many of the
tasks that \cs takes over as ``business logics'' when building the elastic
application on top of OpenStack or Elastic Beanstalk. To give one example, many participants
counted code related to performance monitoring towards ``business logics''.
Note that, due to the open nature of our study tasks, the standard deviations are all
rather large (i.e., solutions using all technologies varied widely in size).
It needs to be noted that the large difference in T1 sizes (for \cs on top
of OpenStack and EC2) between Phase (1)
and Phase (2) solutions can be explained by clarifications in the task descriptions.
In Phase (1), some formulations in the tasks led to much simpler implementations,
while our requirements were formulated much more unambigiously in Phase (2), leading
to more complex (and larger) submissions. Hence, we caution the reader to not compare results
from Phase (1) with those from Phase (2). 

\begin{table}[h!]
\tbl{Time Spent (in Full Hours)}{
\begin{tabular}{rr|c|c|c|c|c|c|c|c}
& &  \multicolumn{4}{c|}{\textbf{Phase (1)}} & \multicolumn{4}{c}{\textbf{Phase (2)}} \\

& &  \multicolumn{2}{c|}{\textbf{JCS/OS}} & \multicolumn{2}{c|}{\textbf{OS}} & \multicolumn{2}{c|}{\textbf{JCS/EC2}} & \multicolumn{2}{c}{\textbf{Beanstalk}} \\
&   & $\tilde{A}$ & $\sigma_{A}$ & $\tilde{B}$ & $\sigma_{B}$ & $\tilde{C}$ & $\sigma_{C}$ & $\tilde{D}$ & $\sigma_{D}$ \\
    \textbf{T1} & & & & & & & & \\
  \hline
&  Tool Learning & \textbf{7} & 2 & 12 & 5 & 28 & 18 & \textbf{16} & 1 \\
&  Coding & \textbf{4} & 10 & 30 & 17 & \textbf{42} & 25 & 54 & 23 \\
&  Bug Fixing & \textbf{7} & 7 & 18 & 12 & \textbf{14} & 8 & 20 & 14 \\
&  Other Activities & \textbf{13} & 9 & 14 & 14 & \textbf{5} & 0 & 6 & 6 \\
&  Entire Application & \textbf{31} & 25 & 76 & 33 & 127 & 25 & \textbf{121} & 36 \\
   \hline
    \textbf{T2} & & & & & & & & \\
  \hline
&  Tool Learning & 8 & 6 & \textbf{2} & 1 & \textbf{15} & 10 & 23 & 18 \\
&  Coding & 30 & 11 & \textbf{25} & 17 & 36 & 5 & \textbf{30} & 14 \\
&  Bug Fixing & 10 & 11 & 10 & 7 & 16 & 16 & \textbf{5} & 0 \\
&  Other Activities & \textbf{7} & 4 & 11 & 10 & \textbf{5} & 0 & 9 & 9 \\
&  Entire Application & 62 & 13 & \textbf{46} & 17 & 125 & 40 & \textbf{102} & 13 \\
\end{tabular}}
\label{tab:hours}
\end{table}



However, looking at lines of code alone is not sufficient to validate our
hypothesis that \cs improves developer productivity, as it would be possible
that the \cs solutions, while being more compact, are also more complicated
(and, hence, take longer to implement). That is why we also asked participants
to report on the time they spent working on their solutions. The results are
compiled in Table~\ref{tab:hours}. We have classified work hours into a number
of different activities: initially learning the technology, coding, testing and
bug fixing, and other activities (e.g., building OpenStack cloud images). Our
results indicate that the initial learning curve for \cs is lower than for
working with OpenStack directly. However, in comparison with Elastic Beanstalk,
some participants reported equal or even more complexity of \cs, mainly because
of the limited tutorial and help information about \cs available in Internet.
For coding \cs appeared to be much faster tool for participants who had at least
some prior experience with cloud computing.

Due to the high standard deviations, looking at this quantitative data alone remains inconclusive. 
Hence, we also analyzed qualitative feedback by the participants in their reports. Multiple developers
have reported that they felt more productive when using \cs. For instance, P01 has stated that
\emph{``the coolest thing about \cs is the reduction of development effort necessary, to host applications in the cloud (\ldots) [there] are a lot of thing you do not have to care about in detail.''}
P03 also concluded that using
\cs \emph{``went a lot smoother than [using OpenStack directly]''}. P07 also seemed to share this sentiment and
stated that \emph{``[After resolving initial problems] the rest of the project was without big problems and I was able to be very productive in coding the solution.''}
In comparison to Elastic Beanstalk, participants indicate that core idea behind \cs is easier to grasp for starting cloud developers than the one behind modern PaaS systems. For example, P13 indicated that \emph{``the [ \cs] API is easier to understand and more intuitive to use. 
Also it fits more into a Java-like programming model, instead of the weird request based approach of the amazon API''}. However, some participants noted that the fact that Elastic Beanstalk is based on common technology also appeals to them. For instance, P10 specified that \emph{``[In case of Elastic Beanstalk,] Well-known technology is the basis for everything (Tomcat/Servlet)''}. Hence, the participant argued that this allows developers who are already familiar with these platforms to be productive sooner.

Summarizing our study results regarding \textbf{RQ1}, we conclude that \cs indeed seems to allow for higher developer productivity than working directly on top of an IaaS API, such as the one of OpenStack. In comparison to AWS Elastic Beanstalk, our results do not clearly indicate more or less productivity with \cs.   

\subsubsection{Comparisong of Developer-Perceived Quality (RQ2)}
\label{sec:subjective}


In order to answer \textbf{RQ2}, we were interested in the participant's subjective evaluation of the used  technologies. Hence, we asked them to rate the technologies along a number of dimensions from 1 (very good) to 5 (insufficient). We report on the dimensions ``simplicity'' (how easy is it to use the tool?), ``debugging'' (how easy is testing and debugging the application?), ``development process'' (does the technology imply an awkward development process?), and ``stability'' (how often do unexpected errors occur?). A summary of our results is shown in Table~\ref{tab:dimensions}.

\begin{table}[h!]
\tbl{Subjective Ratings (lower is better) }{
\begin{tabular}{rr|c|c|c|c|c|c|c|c}
& &  \multicolumn{4}{c|}{\textbf{Phase (1)}} & \multicolumn{4}{c}{\textbf{Phase (2)}} \\
& &  \multicolumn{2}{c|}{\textbf{JCS on OpenStack}} & \multicolumn{2}{c|}{\textbf{OpenStack}} & \multicolumn{2}{c|}{\textbf{JCS on EC2}} & \multicolumn{2}{c}{\textbf{Elastic Beanstalk}} \\
&   & $\tilde{A}$ & $\sigma_{A}$ & $\tilde{B}$ & $\sigma_{B}$ & $\tilde{C}$ & $\sigma_{C}$ & $\tilde{D}$ & $\sigma_{D}$ \\
    \textbf{T1} & & & & & & & & \\
  \hline
&  Simplicity 			& 3 & 0.6 & 3 & 1.2 & 2 & 0 & 2 & 1.4 \\
&  Debugging 			& 3 & 1.5 & 3 & 1 & 4 & 0 & \textbf{3.5} & 0.7 \\
&  Development Process 	& 4 & 1.7 & \textbf{3.5} & 0.5 & \textbf{2} & 1.4 & 3 & 0 \\
&  Stability 			& 2 & 1.4 & 2 & 0.8 & 2 & 1.4 & \textbf{1.5} & 0.7 \\
&  Overall 				& 3 & 0.6 & 3 & 0.8 & 2 & 0 & 2 & 1.4 \\
   \hline
    \textbf{T2} & & & & & & & & \\
  \hline
&  Simplicity 			& \textbf{2} & 0.4 & 3 & 1.4 & \textbf{2} & 1.5 & 3 & 1.4 \\
&  Debugging 			& \textbf{2} & 0.7 & 4 & 1.4 & 4 & 0 & 4 & 0 \\
&  Development Process 	& \textbf{2} & 0.6 & 3 & 1.4 & \textbf{2} & 0 & 2.5 & 0.7 \\
&  Stability 			& 2 & 1.5 & \textbf{1} & 0 & 3 & 0.5 & \textbf{2.5} & 0.7 \\
&  Overall 				& \textbf{2} & 0.4 & 3 & 0 & 3 & 0.5 & 3 & 1.4 \\
\end{tabular}}
\label{tab:dimensions}
\end{table}

Overall, participants rated all used technologies similarly. However, \cs was rated worse than the comparison technologies mainly in terms of ``stability''. This is not a surprise, as \cs still is a research prototype in a relatively early development stage. Participants indeed mentioned multiple stability-related issues in their reports (e.g., P10 mentions that \emph{``When deploying many cloud objects to one host there were behaviors which were hard to reason about''}). Further, some technical implementation decisions in \cs were not appreciated by our study participants. To give an example, P11 noted that \emph{``It is confusing in the configuration that the field AMI-ID actually expects the AMI-Name, not the ID''}.
In contrast, \cs has generally been rated slightly better in terms of simplicity and ease-of use, especially for T2. For example, participant P09 claimed that \emph{``\cs is the clear winner in ease of use. If you quickly want to just throw some Objects in the cloud, it's the clear choice.''}. Similarly, P12 reported \emph{``[\cs is] programmer friendly. All procedure is more low level and as a programmer there are more things to tune and adjust.''}.
In terms of debugging features, all used technologies were not rated overly well. \cs was generally perceived slightly better (due to its local development environment), but realistically all compared systems are currently deemed too hard to debug if something goes wrong.
Finally, in terms of the associated development process, \cs is generally valued highly, with the exception of T1 and \cs on Openstack. We assume that this is a statistical artifact, as the development process of \cs is judged well in all other cases. Concretely P01 stated that with \cs, \emph{``You are able to get application into the cloud really fast. You are not forced to take care about a lot of cloud-specific issues''}.

Independently of the subjective ratings, multiple participants stated that they valued the flexibility that the \cs concept brought over Elastic Beanstalk. Particularly, P11 noted that \emph{``[\cs provides] more flexibility. The developer can decide when to deploy hosts, on which host an object gets deployed, when to destroy a host, etc''}. Additionally, participants favored the monitoring event engine of \cs for performance tracking over the respective features of the PaaS system. For example, P12 specified as an \cs advantage that \emph{``programmatic usage of different events with a powerful event correlation framework [is] in combination with listeners extremely powerful.''}. 

Concluding our discussion regarding \textbf{RQ2}, we note that \cs indeed has some way to go before it is ready for industrial usage. The general concepts of the tool are valued by developers, but currently, technical issues and lack of documentation and technical support make it hard for developers to fully appreciate the power of the \cs model. One aspect that needs more work is how developers define the scaling behavior of their application. Both tasks in our study required the participants to define non-trivial scaling policies, e.g., in order to optimally schedule genetic algorithm executions to cloud resources, which most participants felt unable to do with the current API provided by \cs. Overall, in comparison to working directly on OpenStack, many participants preferred  \cs, but compared to a mature PaaS platform, AWS Elastic Beanstalk still seems slightly preferrable to many. However, it should be noted that \cs still opens up use cases for which using Beanstalk is not an option, for instance for deploying applications in a private or hybrid cloud~\cite{leitner:13}.

\subsection{Runtime Overhead Measurements (RQ3)}

Finally, we investigated whether the improved 
convenience of \cs is paid for with significantly reduced application performance. Therefore, the main goal of these experiments was to compare the performance of the same application built on top of \cs and using IaaS platform (OpenStack or EC2) directly.

\subsubsection{Experiment Setup}

To achieve this, we built a simple sample application (inspired by T2 from the
user study) on top of Amazon EC2 and our private OpenStack cloud. The
application \JSTaaS (``JavaScript Testing-as-a-Service'') provides testing of
JavaScript applications as a cloud service. Clients register with the service,
which triggers \JSTaaS to periodically launch this client's registered test
suites. Results of test runs are stored in a database, which can be queried by
the client. Tests vary widely in the load that they generate on the servers, and
clients are billed according to this load.

Secondly, we also implemented the same application using \cs. As the main goal
was to calculate the overhead introduced by the \cs, we designed both
implementations to have the same behavior and reuse as much business logics code
as possible. In addition, to simplify our setup, focus on execution performance
evaluation and to avoid major platform-dependent side effects, we limited
ourselves to a scenario, where the number of available cloud hosts is static.
The source code of both applications is available
online\footnote{http://www.infosys.tuwien.ac.at/staff/phdschool/rstzab/papers/TOIT14/}.


All four solutions (directly on OpenStack, directly on EC2, and using \cs on both, OpenStack
and EC2) follow a simple master-slave pattern: a single node (the master)
receives tests through a SOAP-based Web Service and schedules them over the set
of available worker nodes. all solutions were tested with a test setup that
consisted of 40 identical parallelizable long-running test suites scheduled
evenly over the set of available cloud machines. Each test suite consisted of a
set of dummy JavaScript tests calculating Fibonacci numbers.
During the evaluation, we measured the total time of an entire invocation of the
service (i.e., how long a test request takes end-to-end, including scheduling,
data transmission, result collection, etc.). A single experiment run consisted
of 10 identical invocations of the testing web service, each time with a
different number of CHs (ranging from 2 to 20 CHs).
To eliminate the influence of other parallel cloud activity, experiment runs for
both versions of evaluation application were running interchangeably for each
number of CHs. To discard accidental results, such experiment
runs were repeated 10 times for each version of evaluation application.

\subsubsection{Experiment Results}

\begin{figure}
\centering
  \includegraphics[width=0.45\linewidth]{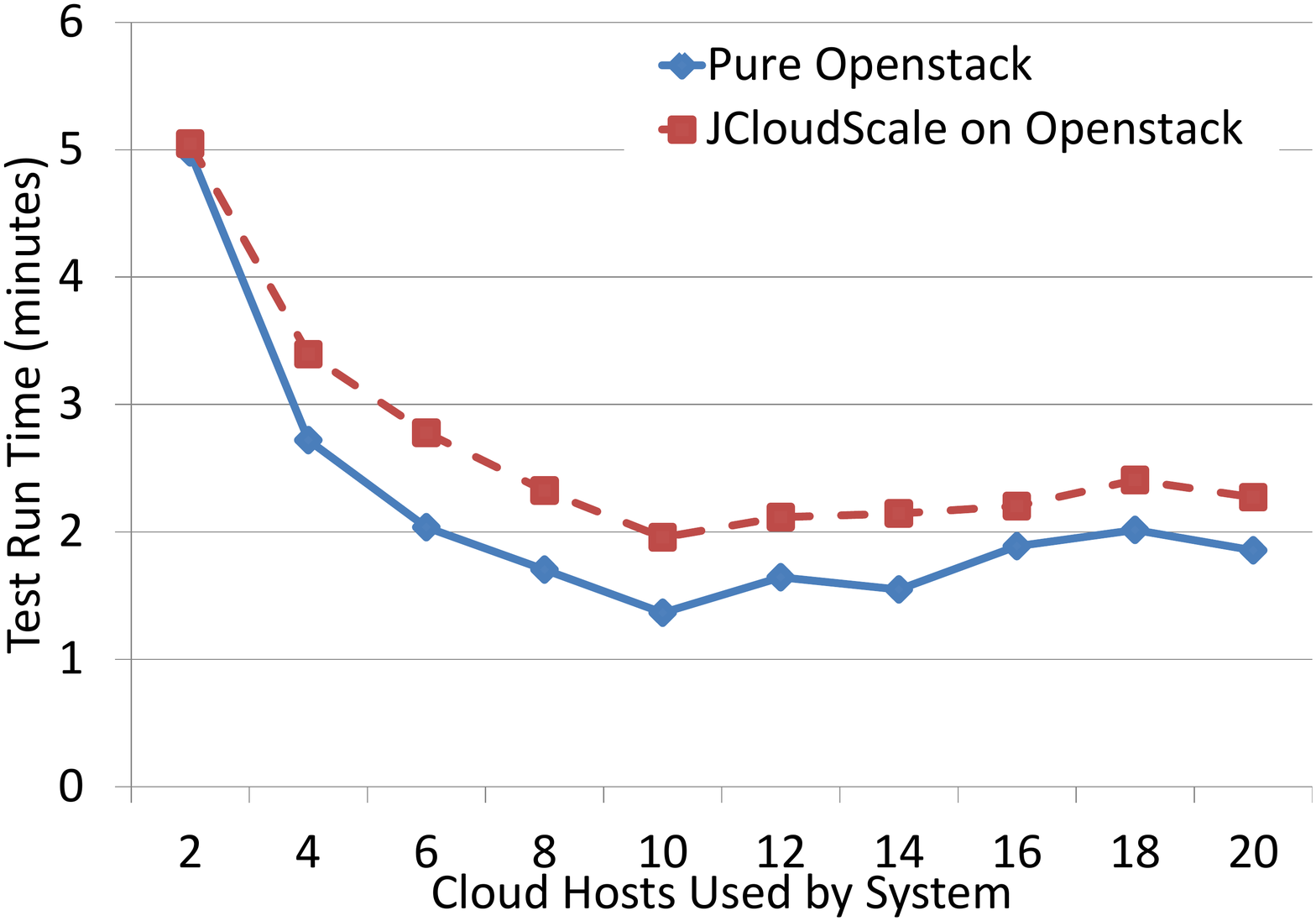}
  \caption{Execution on OpenStack Platform}
  \label{fig:numericEvalOS}
\end{figure}

\begin{figure}
\centering
  \includegraphics[width=0.45\linewidth]{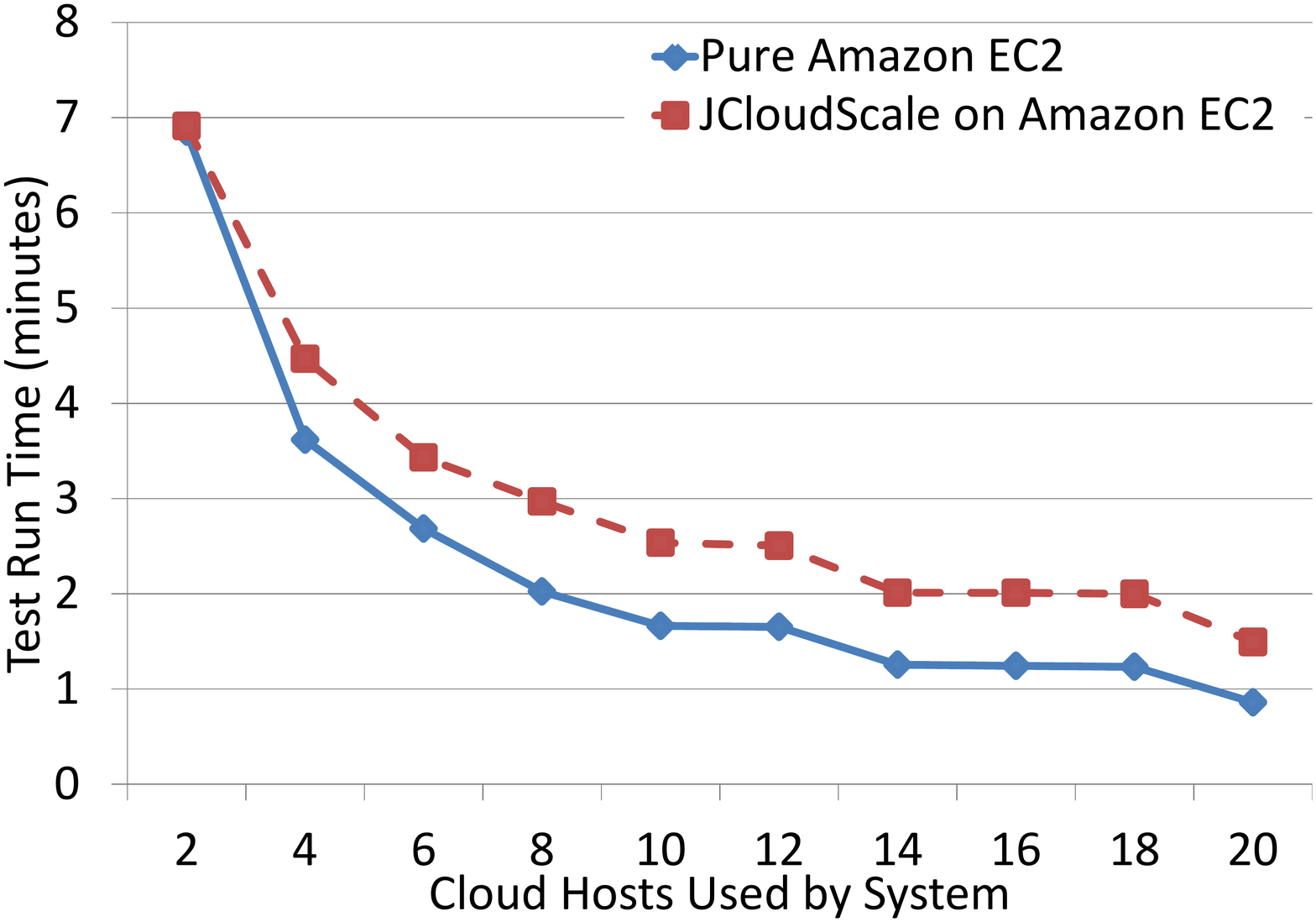}
  \caption{Execution on EC2 Platform}
  \label{fig:numericEvalEC2}
\end{figure}

Figure~\ref{fig:numericEvalOS} and Figure~\ref{fig:numericEvalEC2} show the median total execution time for different numbers of hosts. In general, both applications show similar behavior in each environment, meaning that both approaches are feasible and have similar parallelizing capabilities with minor overhead difference. In both environments, there is an overhead of \cs that is proportional to the amount of used CHs and approximately equal to 2 -- 3 seconds per introduced host for multiple minutes evaluation application. This overhead may be significant for performance-critical production applications, but we believe that it is a reasonable price to pay in the current development stage of the \cs middleware.
However, detailed investigation (and, subsequently, reduction) of the overhead introduced by \cs is planned for future releases of the system.

Another important issue that is visible from Figure~\ref{fig:numericEvalOS} and Figure~\ref{fig:numericEvalEC2} is the cloud performance stability and predictability. With an increasing number of hosts, the total execution time is expected to monotonously decrease, up to a limit when the overhead of parallelization is larger than the gain of having more processors available. This indeed happens in case of Amazon EC2. However, starting with 10 used hosts in OpenStack, the overall application execution time remains almost constant or event increases. In our case, this is mainly caused by 
the limited size of our private cloud. In our system, starting with 10 hosts, physical machines start to get overutilized, and virtual machines start to compete for resources (e.g., CPU or disk IO).

\subsection{Threats to Validity}
\label{sec:threads}

The major thread to (internal) validity, which the results relating to
\textbf{RQ1} and \textbf{RQ2} face, are that the small sample size of 14 study
participants, along with relatively open problem statements, does not allow us
to establish statistical signifiance. However, due to the reports we received
from participants, as well as due to comparing the solutions themselves, we are
convinced that our results that \cs lets developers build cloud applications
more efficiently was not a coincidence. Further, our participants were aware
that \cs is our own system. Hence, there is a chance that our participants
gave their reports a more positive spin, so as to do us a favor. However, given
that all reports contained both negative aspects of all evaluated frameworks,
we are confident that most participants reported truethfully.   
In terms of external validity, it is
possible that the two example projects we chose for our study are not
representative of real-world applications. However, we argue that this is
unlikely, as the projects have specifically been chosen based on real-life
examples that the authors of this paper are aware of or had to build themselves
in the past. Another thread to external validity is that the participants of
our study are all students at TU Vienna. While most of them have some practical
real-life experience in application development, none can be considered
senior developers. 

In terms of \textbf{RQ3}, the major thread to external validity is that the
application we used to measure overhead on is necessarily simplified, and not
guaranteed to be representative of real \cs applications. Real applications,
such as the ones built in our user study, are hard to replicate in exactly the same
way on different systems, hence comparative measurements amongst such systems
are always unfair. To minimize this risk, we have taken care to preserve what we
consider core features of cloud applications even in the simplified measurement
application.

%% file: relatedwork.tex
\section{Related Work}
\label{sec:related}

\begin{sidewaystable}
\scriptsize
\begin{tabular}{p{1.9cm}p{2.9cm}|p{1.5cm}|p{1.5cm}|p{1cm}|p{1.65cm}|p{1.6cm}|p{1.2cm}}
  & & \textbf{Transparent Remoting} & \textbf{Transparent Elasticity} & \textbf{Local Testing} & \textbf{Unrestricted Architecture} & \textbf{Transparent VM Management} & \textbf{Cloud Provider Independence} \\
  \hline
  \multicolumn{2}{c|}{\textbf{Remoting Frameworks}} & & & & & & \\
  Java RMI &  & \centering yes & \centering no & \centering yes & \centering yes & \centering no &  no \\
  Enterprise Java Beans (EJB)  & & \centering yes & \centering partial & \centering yes & \centering yes & \centering no & no \\
  Elastic Remote Methods & \cite{jayaram:13}  & \centering yes & \centering yes & \centering no & \centering yes & \centering yes &  yes \\
  Aneka & \cite{vecchiola:08} \newline \cite{calheiros:12}  & \centering partial & \centering no & \centering no & \centering yes & \centering yes & yes \\
  \hline
  \multicolumn{2}{c|}{\textbf{PaaS Systems}} & & & & & &  \\
  Appengine & & \centering yes & \centering yes & \centering partial & \centering no & \centering yes & no \\
  Amazon Elastic Beanstalk &  & \centering yes & \centering yes & \centering no & \centering no & \centering yes & no \\
  Heroku &  & \centering yes & \centering yes & \centering partial & \centering no & \centering yes & no \\
  AppScale & \cite{chohan:10} \newline \cite{krintz:13}  & \centering yes & \centering yes & \centering no & \centering no & \centering yes & yes \\
  ConPaaS & \cite{pierre:11} \newline \cite{pierre:12}  & \centering yes & \centering yes & \centering no & \centering partial & \centering yes & yes \\
  BOOM & \cite{alvaro:10}  & \centering yes & \centering yes & \centering no & \centering no & \centering no &  yes \\
  Esc & \cite{satzger:11}  & \centering yes & \centering yes & \centering no & \centering no & \centering no & yes \\
  Granules & \cite{pallickara2009}  & \centering yes & \centering yes & \centering yes & \centering no & \centering no &  yes \\  
  \hline
  \multicolumn{2}{c|}{\textbf{Cloud Deployment \& Test Frameworks}} & & &  &  & & \\  
  JClouds &  & \centering no & \centering no & \centering no & \centering yes & \centering no &  yes \\
  Docker &  & \centering no & \centering no & \centering yes & \centering yes & \centering no & yes \\
  Cafe & \cite{mietzner:09} & \centering no & \centering no & \centering yes & \centering no & \centering no & yes \\
  MADCAT & \cite{inzinger:14} & \centering no & \centering no & \centering yes & \centering no & \centering no & yes \\
  OpenTOSCA & \cite{binz:12} \newline \cite{binz:13}  & \centering no & \centering partial & \centering no & \centering yes & \centering yes & yes \\
  \hline
  \hline
  \cs & & \centering yes & \centering  partial & \centering  yes & \centering  yes & \centering  yes & yes \\
    
\end{tabular}
\label{tab:rw}
\end{sidewaystable}

We now put the \cs framework into context of the larger distributed and cloud
computing ecosystem. As the scope of \cs is rather wide, there are a plethora of
existing tools, frameworks and middleware that are related to parts of the
functionality of our system. Based on the descriptions in
Section~\ref{sec:cloudscale} and Section~\ref{sec:elasticity}, we consider the
main dimensions to compare frameworks across are (1) to what extend they
transparently handle remoting and elasticity, (2) how easy it is to locally test and
debug applications, (3) whether the system restricts what kinds of applications can
be built (e.g., only Tomcat-based web applications), (4) whether the system handles
cloud virtual machines for the user, and (5) whether the system is bound to one
specific cloud provider. Systems that are cloud provider independent typically
also can be used in a private cloud context. We provide a high-level comparison
of various systems along these axes in Table~\ref{tab:rw}. We have evaluated
each system along each axis, and assigned ``yes'' (fully supported), ``no'' (no
real support), or ``partial'' (some support). All evaluations are to the
best of the knowledge of the authors of this paper, and based on tool
documentations or publications.

Firstly, \cs can be compared to traditional distributed object
middleware~\cite{emmerich:00}, such as Java RMI or EJB. These systems provide
transparent remoting features, comparable to \cs, but clearly do not provide any
support for cloud specifics, such as VM management. It can be argued that EJB
provides some amount of transparent elasticity, as EJB application containers
can be clustered. However, it is not easy to scale such clusters up and down. A
recent research work~\cite{jayaram:13} has introduced the idea of Elastic Remote
Methods, which extends Java RMI with cloud-specific features.
This work is comparable in goals to our contribution. However, the technical
approach is quite different. Aneka~\cite{vecchiola:08,calheiros:12}, a
well-known .NET based cloud framework, is a special case of a cloud computing
middleware that also exhibits a number of characteristics of a PaaS system. We
argue that Aneka's abstraction of remoting is not perfect, as developers are
still rather intimately aware of the distributed processing that is going on. To
the best of our knowledge, Aneka does not automatically scale systems, and
provides no local testing environment.

Secondly, as already argued in Section~\ref{sec:survey}, many of \cs's features
are comparable to common PaaS systems (Google Appengine, Amazon Elastic
Beanstalk, or Heroku, to name just a few). All of these platforms provide
transparent remoting and elasticity, and take over virtual machine management
from the user. However, all of these systems also require a relatively specific
application architecture (usually a three-tiered web application), and usually
tie the user tightly to one specific cloud provider. Support for local testing
is usually limited, although most providers nowadays have at least limited
tooling or emulators available for download.

In addition to these commercial PaaS systems, there are also multiple platforms
coming out of a research setting. For instance,
AppScale~\cite{chohan:10,krintz:13} is an open-source implementation of the
Google Appengine model. AppScale can also be deployed on any IaaS system,
making it much more vendor-independent than other PaaS platforms. This is
similar to the ConPaaS open source platform~\cite{pierre:11,pierre:12}, which
originates from a European research project of the same name. ConPaaS follows a
more service-oriented style, treating applications as collections of
loosely-coupled servies. This makes ConPaaS suited for a wider variety of
applications, however, it has to be said that ConPaaS still imposes significant
restrictions on the application that is to be hosted.

In scientific literature, there are also a number of PaaS systems which are more
geared towards data processing, e.g., BOOM~\cite{alvaro:10},
Esc~\cite{satzger:11}, or Granules~\cite{pallickara2009}. These systems are hard
to compare with our work, as they generally operate in an entirely different
fashion as compared to \cs or the commercial PaaS operators. However, they
typically only support a very restricted type of (data-driven) application
model, and often do not actually interact with the cloud by themselves. This makes
them necessary cloud provider independent, but also means that developers need to
implement the actual elasticity-related features themselves.

Thirdly, we need to compare \cs to a number of cloud computing related frameworks, which
cover a part of the functionality provided by our middleware. JClouds is a Java library
that abstracts from the heterogenious APIs of different IaaS providers, and allows
to decouple Java applications from the IaaS system that they operate in. \cs internally
uses JClouds to interact with providers. However, by itself, JClouds does not provide
any actual elasticity. Docker is a container framework geared towards bringing
testability to cloud computing. Essentially, Docker has similar goals to the local
test environment of \cs.

\cs also has some relation to the various cloud deployment models and systems
that have recently been proposed in literature, e.g., Cafe~\cite{mietzner:09},
MADCAT~\cite{inzinger:14}, or OpenTOSCA~\cite{binz:12,binz:13}, which is an open
source implementation of an upcoming OASIS standard. These systems do not
typically cover elasticity by themselves (although TOSCA has partial support for
auto-scaling groups), but they are usually independent of any concrete cloud
provider.

By design, \cs supports most of the characteristics we discuss here. However,
especially in comparison to PaaS systems, developers of \cs applications are not
entirely shielded from issues of scalability. Further, as the user study
discussed in Section~\ref{sec:survey} has shown, the system still needs to
improve how scaling policies are written, so as to make building elastic systems
easier for developers.

%% file: conclusions.tex
\section{Conclusions}
\label{sec:conclusions}

\cs is a Java-based middleware that eases the development of elastic cloud
applications on top of an IaaS cloud. \cs follows a declarative approach based
on Java annotations, which removes the need to actually adapt the business
logics of the target application to use the middleware. Hence, \cs support can
easily be turned on and off for an application, leading to a flexible
development process that clearly separates the implementation of target
application business logics from implementing and tuning the scaling behavior.

We have introduced the core concepts behind \cs, and presented an
evaluation of the middleware based on an user study as well as using a simple
case study application. Our results indicate that \cs is well received among
initial developers. Our results support our claim that the general \cs model
has advantages to both, working directly on top of an IaaS API or on an industrial
PaaS systems. However, further study is
required to strengthen these claims, as the limited scale of our initial study
was not sufficient to clear all doubts about the viability of the system.
Further, there are also technical and conceptual issues that require further
investigation. Most importantly, we have learned that implementing actually
elastic applications is still cumbersome for developers, as getting the scaling
policy right is not as easy as we had hoped.

%
%